# Excitonic parameters of GaN studied by time-of-flight spectroscopy

T. V. Shubina[1]*, A. A. Toropov[1], G. Pozina[2], J. P. Bergman[2], M. M. Glazov[1],

N. A. Gippius[3]**, P. Disseix[3], J. Leymarie[3], B. Gil[4], and B. Monemar[2]

[1]*Ioffe Physical-Technical Institute, RAS, St Petersburg 194021, Russia*

[2]*Department of Physics, Chemistry and Biology, Linköping University, S-581 83 Linköping, Sweden*

[3]*LASMEA-UMR 6602 CNRS-UBP, 63177 AUBIERE CEDEX, France*

[4]*GES, UMR5650, Université Montpellier 2–CNRS, Montpellier, France*

We refine excitonic parameters of bulk GaN by means of time-of-flight spectroscopy of light pulses propagating through crystals. The influence of elastic photon scattering is excluded by using the multiple reflections of the pulses from crystal boundaries. The shapes of these reflexes in the time-energy plane depict the variation of the group velocity induced by excitonic resonances. Modeling of the shapes, as well as other spectra, shows that a homogeneous width of the order of 10 µeV characterizes the exciton-polariton resonances within the crystal. The oscillator strength of A and B exciton-polaritons is determined as 0.0022 and 0.0016, respectively.

The current interest in the phenomenon of slow light is mostly related to the fact that any realization of all-optical schemes of quantum communication and information processing implies a deterministic retardation of photon pulses [1]. The phenomenon of light delay due to optical dispersion induced by a resonance line [2] may be important for a set of optoelectronic devices, where light should propagate a long distance, or for those operating at a high-frequency. Here, we demonstrate that the study of the slow light may bring another benefit. It allows one to refine excitonic parameters in semiconductors.

The light propagation through a medium in the vicinity of the resonance may be either ballistic (polariton-like), with conservation of wave vector [3], or diffusive, when the wave vector, and possibly frequency, is changed during acts of scattering [4]. Recently it has been demonstrated that both mechanisms can coexist in GaN, where the light is scattered by neutral donor bound exciton ($D^0X$) complexes [5]. As a result, the transmitted optical signal is an admixture of both ballistic and diffusive components that hampers the correct modeling of experimental dependences. Fortunately, the time-of-flight spectroscopy permits one to record weak reflexes in the time-resolved (TR) images of a transmitted signal. These reflexes, emerging due to the internal reflection of light pulses from the sample boundaries, were recorded in GaN [5] and later in CdZnTe crystals [6]. The mechanism of their propagation is exclusively ballistic, because otherwise photons lose any memory about the initial direction after a few acts of scattering. The delay and shape of these reflexes reproduce the variation of the group velocity $v_g(\omega)$, which, in turn, is controlled by the parameters of the excitonic resonances.

For GaN, there is a huge variation in reported values of the exciton oscillator strength (*f*) and the directly linked quantity longitudinal-transverse splitting ($\omega_{LT}$). This uncertainty is determined by various factors, such as elastic strain in the sample [7], whose exact value is frequently unknown, different quality of samples, including the tilt of columns [8], and certainly experimental methods. As an example, an extraordinarily high value of $\omega_{LT}$ can be reported, when it was defined as a separation between photoluminescence (PL) peaks, because a thermalization process can significantly enhance apparent gap [9]. The commonly used modeling of reflectivity spectra [10,11] suffers from uncertainty caused by a large number of the fitting parameters (up to 16 for one polarization). Besides, both PL and reflectivity techniques probe an area close to the sample surface, rather than its volume, whereas a certain discrepancy between the properties of these regions may arise due to different strain, contamination, and local electric field-induced bands modification [12]. A similar uncertainty exists for the width of the resonances, whose reported values vary from 0.2 meV [10] up to 2.1 meV [13] for A exciton. This parameter is commonly considered as an empirical damping constant ($\Gamma_{eff}$). However, the intrinsic linewidth has a certain physical meaning. It characterizes the decay of the coherence for the localized excitation or the dephasing relaxation for exciton-polaritons [14]. Further progress in development of smart GaN-based devices, like polaritonic lasers [15], requires knowledge of these important characteristics.

In this paper, we propose to use the data on the reflexes of the light pulses to refine the excitonic parameters of bulk GaN. The simultaneous fitting of both time-resolved images and continuous wave (cw) transmission spectra, taking into account reflectivity data, gives us an opportunity of accurate determination of the exciton-polariton parameters. Our data evidence that the homogeneous linewidth of the resonances in bulk, being of the order of ten μeV, has nothing in common with the effective damping constant commonly used to fit the reflectivity spectra. The oscillator strength is found to be smaller than most of the reported values, determined by other methods. Besides, we show that the study of the light propagation is a unique way to estimate the parameters of $D^0X$ resonances, which can hardly be done by other methods.

The samples under investigation were free-standing GaN layers with thickness $L$ of 1 and 2 mm grown by hydride vapor phase epitaxy (HVPE), which are currently among the best available. This is confirmed by extremely narrow PL lines of the bound excitons (spectral width of 0.3 meV) and pronounced PL peaks ascribed to the exciton-polariton emission [16]. The TR experiments were done at 2 K, using the pulses of a tunable picosecond laser (Mira-HP) and a Hamamatsu streak camera with spectral resolution about 2 ps. The dynamical range of the detecting system was not less than $10^3$. The transmission and reflectivity spectra were measured at 5 K with the normal light incidence using a tungsten lamp.

As illustrated by the scheme in Fig. 1, the pulse reflexes can be registered in two experimental configurations. In the back-scattering configuration, the even reflexes (2nd and 4th) propagate in the sample a distance of 2$L$ and 4$L$, respectively, while in the transmission geometry the odd 3rd reflex propagates 3$L$, *etc*. In this study, we focus on the even-order reflexes (Fig. 2), because their delay is well-defined with respect to an initial pulse (denoted as 0$L$), directly reflected from the sample surface. Figure 1 (a) shows the spectral dependencies of the delay times measured along the pulses of different energies for the same length of propagation, namely, through the 2-mm-long sample and for 2$L$ replicas in the 1-mm-long sample. The delay time of the basic transmitted pulse in the 1-mm sample, multiplied by two in order to simulate the 2-mm length, is shown for comparison as well. Obviously, the delay times are smallest for the 2$L$ reflex, when the photon scattering, enhancing the light path, is excluded. The difference in the dependences for the 1-mm and 2-mm samples is related to the different donor concentration ($5 \cdot 10^{15}$ cm$^{-3}$ vs $10^{17}$ cm$^{-3}$, respectively), which influences the scattering cross-sections [5]. Figure 2 shows the sequence of the appearance of the 2$L$ and 4$L$ reflexes arising with the shift of the pulse from the D$^0$X lines towards a transparency region. Near the D$^0$X lines, these reflexes are strongly curved and their higher-energy edge is extended along the time axis until full disappearance.

To analyze the experimental data we use the model of the ballistic light propagation in a medium with several resonances, which, in general, can be inhomogenously broadened [7]. The delay $T(\omega) = L/v_g(\omega)$ is given by the group velocity $v_g(\omega) = d\omega/dk$, where the wave vector $k(\omega) = (\omega/c)\sqrt{\varepsilon(\omega)}$. The complex dielectric function $\varepsilon(\omega)$ is written as:

$$\varepsilon(\omega) = \varepsilon_b + \sum_j \int \frac{f_j \omega_{0,j}}{\omega_{0,j} + \beta k^2 + \xi - i\Gamma_j - \omega} \frac{1}{\sqrt{\pi}\Delta_j} \exp(\frac{-\xi^2}{\Delta_j^2}) d\xi. \qquad (1)$$

Here, $\varepsilon_b = 9.5$ is the background dielectric constant. Each $j$-term within the sum is the convolution of the resonance line, characterized by the homogeneous width $\Gamma$, with the Gaussian centered on the same frequency $\omega_0$. The parameter $\Delta_j$ describes the inhomogeneous width of the broadened resonances. Spatial dispersion is taken into account by the term $\beta k^2 = (\hbar^2 / 2M_j) k^2$, where the effective masses $M_j$ equal $0.5m_0$, $0.6m_0$, and $0.8m_0$ ($m_0$ - electron mass) for the A, B, and C excitons, respectively [10], being considered as infinite for $D^0X$. Such a representation is allowed for frequencies not too close to $\omega_0$ of the free exciton resonance that is sufficient for the consideration of our data on cw and TR light transmission. The estimations showed that the changes made by the spatial dispersion are not essential, being within the 20% limit. Therefore, Eq. (1) with the omitted $\beta k^2$ term can be used to simulate the reflectivity spectra using the equation:

$$R(\omega) = |[(n(\omega) - 1)^2 + \kappa(\omega)^2] / [(n(\omega) + 1)^2 + \kappa(\omega)^2]|, \qquad (2)$$

where $n(\omega) = \mathrm{Re}\sqrt{\varepsilon(\omega)}$ and $\kappa(\omega) = \mathrm{Im}\sqrt{\varepsilon(\omega)}$. The intensity of the transmitted signal is estimated as:

$$I(\omega) = (1 - R)^2 \exp(-D) / [1 - R^2 \exp(-2D)], \qquad (3)$$

where $D = L\alpha$ is the optical density, $\alpha = 2\omega\kappa(\omega)/c$ is the absorption coefficient.

The fitting procedure includes the simulation of reflectivity spectra [Fig. 1 (b)] to determine approximately the energies $E_j = \hbar\omega_{0,j}$ and $f$ values of the exciton-polariton resonances. At this stage, the inhomogeneous width $\Delta_j$ to some extent plays the role of an effective damping parameter, used with the conventional modeling of reflectivity spectra. We underline that the determined parameters, $\Delta_A = 0.85$ and $\Delta_B = 1.0$ meV, characterize rather a region close to surface, than the bulk GaN. Next step is the simulation of the cw transmission spectra [Fig. 1 (c)], that gives us the value of the homogeneous width $\Gamma$ of the exciton-polariton resonances. This magnitude refers to the GaN volume, where the pulses propagate. Absorption of light in a crystal depends on the imaginary part of the dielectric function, which is predominantly determined by this term. The final correction of the excitonic parameters is made by means of the simulation of the delay, curvature, and attenuation of the transmitted reflexes. The particular delay of the pulse constituents which provides its curvature along the time axis depends predominantly on the oscillator strength. Note that such a dominant influence of each of the parameters on a certain process simplifies the numerical simulation.

This simulation shows that the influence of the $D^0X$ resonances is important on small energy scale; it is noticeable only at a distance less than 1-2 meV from the PL peaks [see Fig. 2 (d)]. Consideration of the reflexes outside this range gives us the parameters of the exciton polaritons. In this modeling, we assumed that the parameters of the $D^0X$ resonances are similar for both lines $O^0X_A$ (3.4714 eV) and $Si^0X_A$ (3.4723 eV), related to the oxygen and silicon donors. Their oscillator strength and homogeneous width for the 1-mm sample are defined as $5 \cdot 10^{-6}$ and 1 µeV, respectively (it may be different for other donor concentrations). A deviation from these values exceeding 10% does not allow one to reproduce the shape of the reflexes, shown in Fig. 2 (a-c). Note that the $f_{D^0X}$ value, derived from the fitting of the reflexes, is 4-5 times less than the value estimated using the basic transmitted pulse, assuming significant contribution of the photon diffusion [5]. The inhomogeneous width of the $D^0X$ resonances is determined by the consideration of the light attenuation near these lines. The excess over the 35 µeV causes quenching of the higher-energy tail of the light reflexes at the energies, where the signal is observed in the experimental images. It provides the shift of the apparent pulse center towards lower energy, as it is illustrated by the inset in Fig. 2 (a). The obtained values for the $D^0X$ resonances are substituted into Eq. (1) to determine finally the parameters of the exciton-polaritons, inherent for the bulk GaN. They are shown in Table 1 together with the most frequently cited previously reported data.

Table 1. Summary of exciton-polariton parameters of bulk hexagonal GaN derived from cw and TR transmission data, shown together with reported data obtained by a reflectivity technique ($\Gamma_{eff}$ is the effective damping parameter used in the reflectivity modeling). The accuracy is $\pm 0.3$ meV for the energies, being within $\pm 10\%$ limits for other parameters.

| Sample | Parameter | A | B | C | Ref. |
|---|---|---|---|---|---|
| HVPE free standing crystals | E, meV | 3478.4 | 3483.4 | 3501.6 | This work |
| | f | 0.0022 | 0.0016 | 0.0004 | |
| | $\Delta$, meV | 0.85 | 1.0 | 1.0 | |
| | $\Gamma$, µeV | 13 | 13 | 13 | |
| LEO method grown layers | E, meV | 3479.1 | 3484.4 | 3502.7 | [10] |
| | f | 0.0033 | 0.0029 | 0.0007 | |
| | $\Gamma_{eff}$, meV | 0.2 | 0.7 | 1.2 | |
| Homoepitaxial MOCVD layers | E, meV | 3476.7 | 3481.5 | 3498.6 | [11] |
| | f | 0.0027 | 0.0031 | 0.0011 | |
| | $\Gamma_{eff}$, meV | 0.7 | 1.5 | 3.1 | |

The oscillator strength values, being about 0.0022 and 0.0016 for the A and B excitons, respectively, are smaller than the values determined using the surface-probing reflectivity data [10,11]. In the region below $\omega_{0,A}$, the A exciton resonance affects the light propagation stronger than remote B and, especially, C excitons. Therefore, the accuracy of its parameter determination is high. Note that further increase of the *f* value for the B exciton would unrealistically diminish the A-exciton *f* value. The longitudinal-transverse splitting for the A excitonic series is 0.8±0.1 meV, estimated as $\omega_{LT} = f_j \omega_{0,j} / \varepsilon_b$. This value is somewhat less than the splitting ~1 meV between the PL peaks, ascribed to transverse and longitudinal A exciton emission using polarization-dependent PL measurements in the same 1-mm sample [16]. The homogeneous width of 13±1 µeV, assumed to be the same for all exciton-polariton resonances, is similar to the value recorded for polaritons in other direct gap semiconductor CuCl (3.4 eV room-temperature bandgap) [14]. At low temperatures, this parameter has to be small in perfect crystals, being determined mostly by the acoustic phonon scattering [17]. The inhomogeneous widths of all exciton-polariton resonances do not influence markedly the light reflexes, which are observed below the $D^0X$ lines. Thus, the correct values of $\Delta_j$ cannot be derived from these experiments. There is one limitation, namely, the $\Delta_j$ value should be less than $\omega_{LT}$ to allow the survival of exciton-polaritons [18]. Thus, we cannot exclude that the exciton-polaritons propagate at low temperatures through the GaN crystal as coherent excitations with narrow homogeneous linewidth.

In conclusions, the parameters of excitonic resonances in GaN are determined using time-of-flight spectroscopy studies, done in combination with reflectivity and cw transmission measurements. Consideration of the slow light reflections from the crystal boundaries allows us to keep out the influence of the resonant photon scattering. Our data demonstrate that the exciton-polaritons propagating within a GaN crystal have homogeneous width of the order of 10 µeV and the oscillator strength, which is 20-30% lower than the currently accepted value.

We thank A. Vasson for the help in these studies. This work has been supported in part by the RFBR (grant 10-02-00633), the Program of the Presidium of RAN, and the Dynasty Foundation. TVS acknowledges the Université Montpellier 2 for its hospitality.
[1] M. O. Scully and M. S. Zubairy, Science **301**, 181 (2003).

[2] R. Loudon, J. Phys. A **3**, 233 (1970).

[3] R. G. Ulbrich and G. W. Fenrenbach, Phys. Rev. Lett. **43**, 963 (1979).

[4] T. Steiner, M. L. W. Thewalt, E. S. Koteles, and J. P. Salerno, Phys. Rev. B **34**, 1006 (1986).

[5] T. V. Shubina, M. M. Glazov, A. A. Toropov, N. A. Gippius, A. Vasson, J. Leymarie, A. Kavokin, A. Usui, J. P. Bergman, G. Pozina, and B. Monemar, Phys. Rev. Lett. **100**, 087402 (2008).

[6] T. Godde, I. A. Akimov, D. R. Yakovlev, H. Mariette, and M. Bayer
Phys. Rev. B **82**, 115332 (2010).

[7] B. Gil, F. Hamdani, and H. Morkoç, Phys. Rev. B **54**, 7678 (1996).

[8] D. C. Reynolds, B. Jogai, and T. C. Collins, Appl. Phys. Lett. **80**, 3928 (2002).

[9] B. Gil, A. Hoffmann, S. Clur, L. Eckey, O. Briot, and R.-L. Aulombard, J. Crystal Growth **189/190**, 639 (1998).

[10] K. Torii, T. Deguchi, T. Sota, K. Suzuki, S. Chichibu, and S. Nakamura, Phys. Rev. B **60**, 4723 (1999).

[11] R. Stepniewski, K. P. Korona, A. Wysmolek, J. M. Baranovski, K. Pakula, M. Potemski, G. Martinez, I. Grzegory, and S. Porowski, Phys. Rev. B **56**, 15 151 (1997).

[12] B. Monemar, P. P. Paskov, J. P. Bergman, G. Pozina, A. A. Toropov, T. V. Shubina, T. Malinauskas, and A. Usui, Phys. Rev. B **82**, 235202 (2010).

[13] Y. Toda, S. Adachi, Y. Abe, K. Hoshino, and Y. Arakawa, Phys. Rev. B **71**, 195315 (2005).

[14] T. Takagahara, Phys. Rev. B **31**, 8171 (1985).

[15] S. Christopoulos, G. Baldassarri H. von Högersthal, A. J. D. Grundy, P. G. Lagoudakis, A.V. Kavokin, J. J. Baumberg, G. Christmann, R. Butté, E. Feltin, J.-F. Carlin, and N. Grandjean, Phys. Rev. Lett. **98,** 126405 (2007).

[16] B. Monemar, P. P. Paskov, J. P. Bergman, A. A. Toropov, T. V. Shubina, T. Malinauskas, and A. Usui, phys. stat. sol. (b) 245, 1723 (2008).

[17] B. Segall and G. D. Mahan, Phys. Rev. **171**, 935 (1968).

[18] Y. Masumoto, S. Shionoya, and T. Takagahara, Phys. Rev. Lett. **51**, 923 (1983).

[19] M. Matsushita, J. Wicksted, and H. Z. Cummins, Phys. Rev. B **29**, 3362 (1984).

List of figure captions

Fig. 1. (color online) (a) Delay time dependencies obtained using (1) – light pulse reflexes in the 1-mm sample; (2) – basic pulses transmitted through the same sample, multiplied by two; (3) – basic pulses in the 2-mm sample. The inset shows the scheme of the back-scattering and transmission measurements, where n·$L$ denotes the series of replicas. (b) Experimental spectra

of reflectivity and (c) transmission, measured in the 1-mm sample. The thin red lines are simulated spectra calculated as described in the text.

Fig. 2. (color online) Selected TR images of replicas of impinging pulses at the different energies, recorded in the back-scattering configuration. The delay time dependencies calculated for the 2$L$ and 4$L$ reflexes (black lines) are plotted over these images. The red line in (d) is the delay time dependence calculated neglecting the $D^0X$ resonances. The inset shows the contour of the transmitted pulse with $\Delta_{D^0X}$ assumed (1) – 0.035 meV and (2) – 0.1 meV.

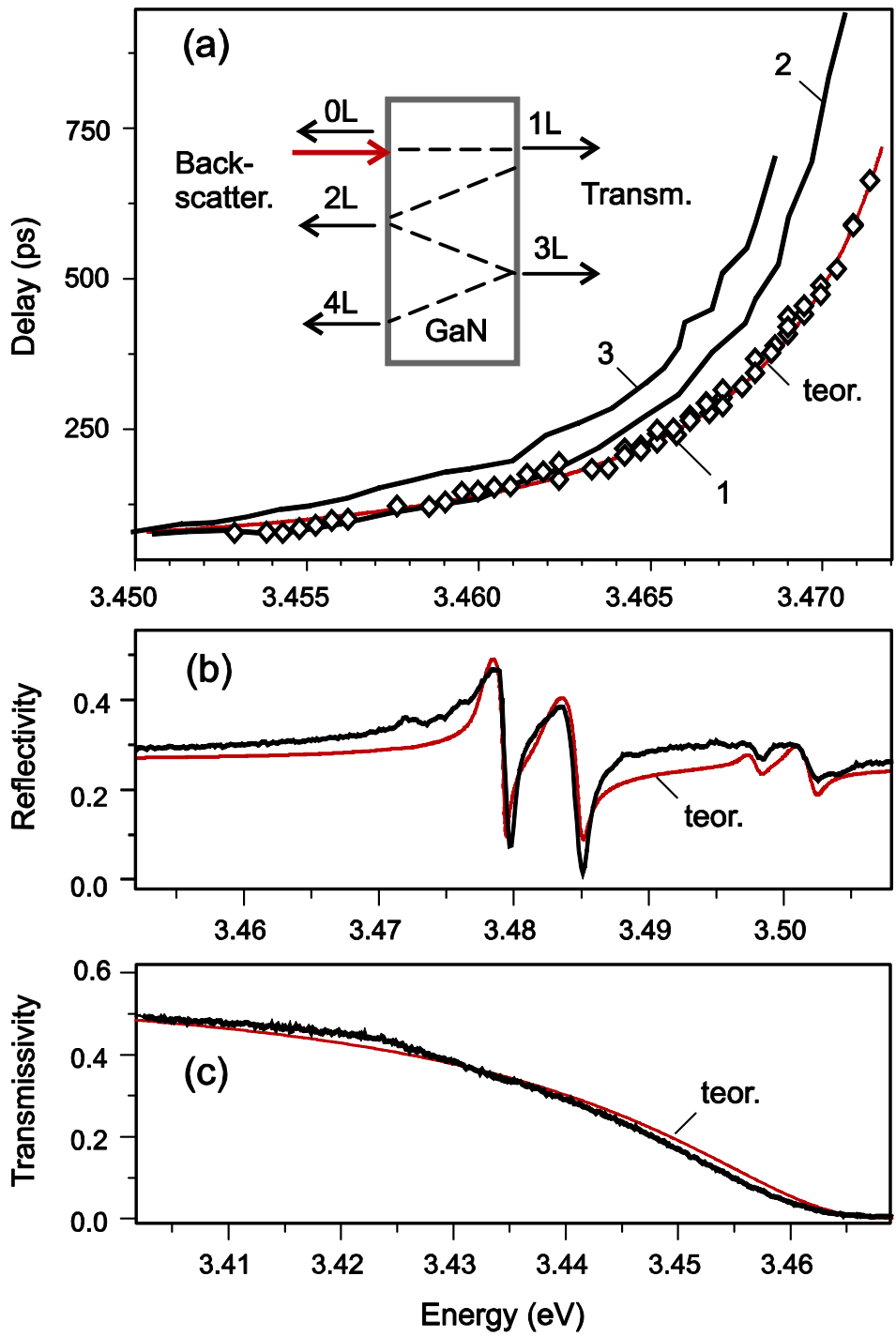

Fig. 1

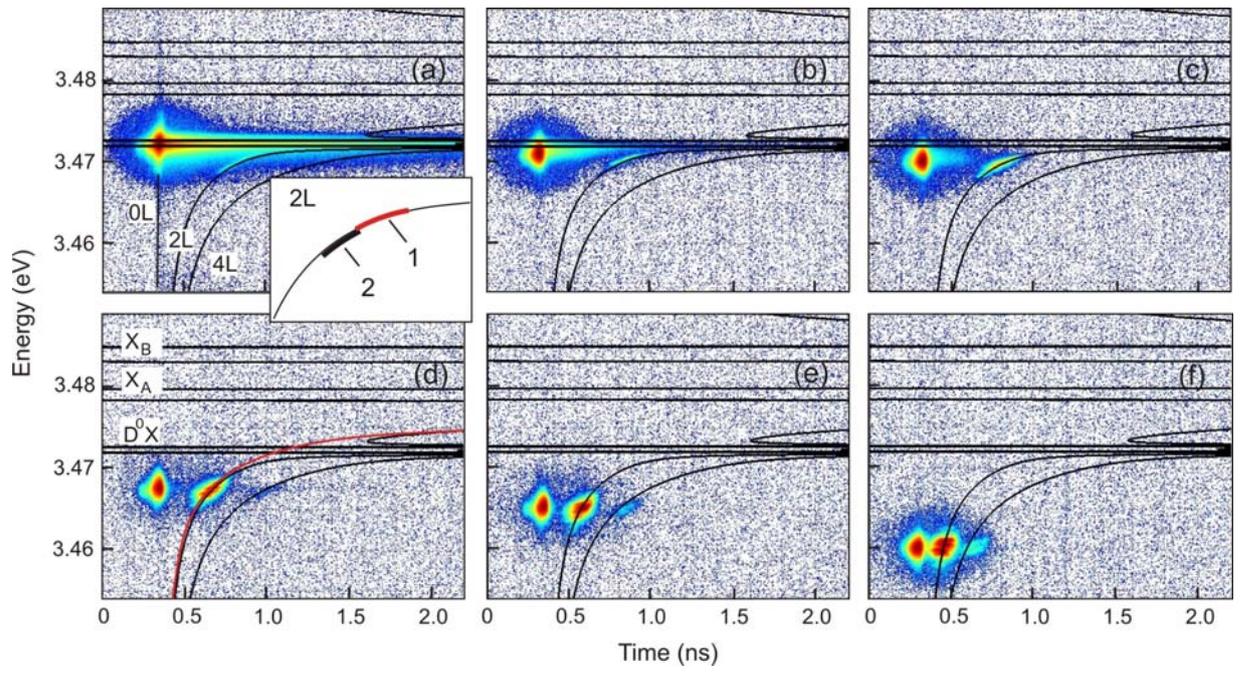

Fig. 2